# A Hierarchical Coding-Window Model of Parkinson's Disease

Andres Daniela Sabrina[1,2,3,*], Gomez Florian[1], Cerquetti Daniel[2], Merello Marcelo[2], Stoop Ruedi[1]

[1] Institute of Neuroinformatics, ETH and UZH Zurich, Winterthurerstrasse 190, 8057, Zurich, Switzerland
[2] Institute for Neurological Research Raul Carrea, Fleni Institute, Movement Disorders Section, Buenos Aires, Argentine
[3] Society in Science, The Branco-Weiss Fellowship, administered by ETH, Zurich, Switzerland

* Corresponding author: dandres@ini.uzh.ch

**Abstract.** Parkinson's disease is an ongoing challenge to theoretical neuroscience and to medical treatment. During the evolution of the disease, neurodegeneration leads to physiological and anatomical changes that affect the neuronal discharge of the Basal Ganglia to an extent that impairs normal behavioral patterns. To investigate this problem, single Globus Pallidus pars interna (GPi) neurons of the 6-OHDA rat model of Parkinson's disease were extracellularly recorded at different degrees of alertness and compared to non-Parkinson control neurons. A structure function analysis of these data revealed that the temporal range of rate-coded information in GPi was substantially reduced in the Parkinson animal-model, suggesting that a dominance of small neighborhood dynamics could be the hallmark of Parkinson's disease. A mathematical-model of the GPi circuit, where the small neighborhood coupling is expressed in terms of a diffusion constant, corroborates this interpretation.

**Keywords:** Parkinson's disease, neuronal code, structure function, diffusive coupling

For the coding of neuronal information, the precise relative timing of neuronal firing within the neuronal ensemble is an attractive scheme. However, also the average discharge rate of neurons or of neuronal ensembles has its advantage: clearly, such a code would be more robust, in particular with respect to noise that we find ubiquitous in the cortex [1, 2]. The disadvantage of rate coding is a smaller maximal information content and an increased latency. The discharge time-window across which a rate is calculated has necessarily two limiting time scales: a time beyond which further spikes no longer modify the rate-coded information (long-term correlations, however, might be present nonetheless) and a short time scale given by the finite resolution beyond which two spikes can no longer be distinguished from a single one. As different environmental conditions will induce different information scales on the neural system, it can be expected that different coding systems co-exist, but have their distinguished relevance for different time-scales.

Parkinson's disease (PD) is characterized by a conflict between different intrinsic neuronal firing time scales [3]. To distinguish between characteristic time scales, structure functions are a well-tested approach [4-6]. Comparing structure functions with autocorrelation analysis, both offer essentially the same information, but structure functions have been shown to be more robust to the presence of drift, low frequency noise and short time series [7]. Also, the advantage of studying a family of structure functions is that geometrical properties of the neuronal discharge could be captured in terms of corresponding scaling exponents [8]. We apply the structure function approach to interspike interval (ISI) time series, where $I(j)$ is the $j$th interspike interval and $\Delta I(\tau) = I(j+\tau) - I(j)$ is the difference between successive intervals, separated by an index increment $\tau \in N+$. The structure function $S_q(\tau)$ is defined as

$$S_q(\tau) = \left\langle \left| \Delta I(\tau)^q \right| \right\rangle, \qquad (1)$$

where $\langle \cdot \rangle$ accounts for the statistical average over the time series and $q$ is a real number. The scaling behavior of $S_q$ is then characterized by the power law relationship

$$S_q(\tau) \sim \tau^{\zeta(q)}. \qquad (2)$$

For a stationary process with independent increments $\zeta(q) = 0$, which expresses that the mean correlation between successive events does not depend on the event index [9]. Monofractal, non-intermittent time series imply $\zeta(q) = const$, whereas multifractal behavior is characterized by $\zeta(q)'' < 0$. The zero-slope regime ($\zeta(q) = 0$) of the structure function is of particular interest, since it marks the temporal scale across which only random processes are at work. For neuronal signals, this regime precludes coding schemes other than a rate code.

We applied the structure function analysis to data from single neurons of the Globus Pallidus pars interna (GPi). Spontaneous neuronal firing was recorded in an animal model of PD (6-hydroxydopamine - 6OHDA - partial lesion model of Sprague-Dawley adult rats) and a control group. Experiments were performed at two conditions, firstly with the animals under deep chloral-hydrate anesthesia and following at full alertness, both in relaxed, head-restrained conditions. We analyzed 52 neuronal recordings in total, 27 of the PD group (13 under anesthesia and 14 in the awake state) and 25 of the control group (13 under anesthesia, 12 awake). Details of the procedures of the animal experiments and recording technique can be found in the supplementary material (see Appendix). Signals were processed off-line; spikes were extracted and classified using standard spike-sorting [10]. Raw neuronal recordings were around 60 minutes long and the time series extracted included roughly 10000 events (14731 ± 7210 ISI, mean ± SD). Interspike intervals (ISI) time series were constructed and structure functions of increasing order were calculated for each time series. For the structure function analysis, we varied $\tau$ between 1 and 2000. For simplicity, q was restricted to the set $\{1,2,...,6\}$. The behavior of the structure functions was checked to be consistent for $1 \leq q \leq 6$ and since they show a low dependence on order $q \leq 6$, the detailed analysis will be restricted to order one.

In the majority of the neurons studied, three distinct regimes of the temporal structure function emerge (separated by index-points $s_1$ and $s_2$, cf. Fig. 1). Roughly, breakpoint $s_1$ is defined as the locus where the initially ascending behavior changes abruptly into an essentially constant state, the end of which is indicated by breakpoint $s_2$. More precisely, for the location of the breakpoints, $\log(S_q(\tau))$ was plotted vs. $\log(\tau)$ and a first-order function was fitted to it. The main part of regime II was obtained by maximizing the range over which the fitted function would have absolute slope less than exp (-3), with data scattered around it having a standard error of the same size. This area was then extended to meet with the monotonous behavior manifestly present at smaller and larger scales. Occasionally, instead of ascending regimes I / III, we observed descending behavior. Such variants were also encountered during our model simulations, and thus appear to describe a natural phenomenon (see below). In only 7 out of the 52 cases studied (13%), regime I presented a descending instead of an ascending behavior. This means that there aren't any small-scale processes with memory affecting the time series at a scale larger than unity (ISI). In the case of the long-term correlations regime III, an ascending behavior was found in 40 out of 52 neurons studied (77%). This regime reflects the presence of memory processes at long time-scales. However, its study is limited by the finite length of the time series. Therefore, a descending behavior of regime III could reflect the presence of higher-scale oscillations, and the distinction between the implications of an ascending or a descending behavior of regime III is less clear.

Functionally, regime II indicates the window where temporal rate-coding is expected. This regime is substantially reduced in the PD group at all alertness levels. In 9 out of 27 recorded PD neurons, a shortened regime II was found, while in the remaining cases regime II had fully disappeared (see for example Fig. 1, panel B). The interpretation that the disappeared windows are extreme cases of shortened windows was corroborated by a subsequent simulation study (see below). To translate our findings from the index space underlying the structure function to the time-axis, we multiplied the indices $s_1$ and $s_2$ by the average ISI duration of the time series, from which we obtained the corresponding characteristic time-breakpoints $t_1$ and $t_2$. Using time as the abscissa, under anesthesia, PD neurons show a substantially increased $t_1$ (on average 38 sec. vs. 1 sec. on average, p=0.07, two sample t-test). At full alertness, $t_2$ is more than eight times smaller in PD neurons compared to the control group (4 sec. vs. 35 sec. on average, p=0.01). Regarding the anaesthesia-alertness transition, whereas for the control group it is characterized by the increase of $t_2$ (from 8 to 35 sec., p=0.01), the extension of the long-range correlation regime III is the main feature in the PD group.

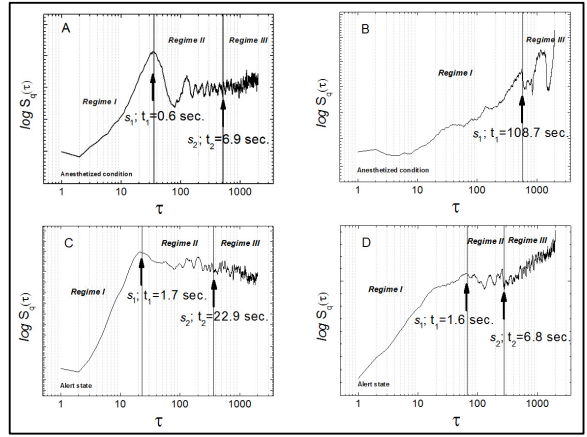

**Fig. 1) Order-1 structure functions of the four experimental groups** (characteristic examples). A), B) Deep anesthesia condition, control and PD neuron, respectively. C), D) Full alertness condition, control and PD neuron, respectively. Smoothing over 20 data points was applied to the original data. Arrows: times $t_1$ and $t_2$ corresponding to breakpoints *s1* and *s2*. Control neurons show a clear zero-slope scale-range (regime II) at both alertness conditions, prolonged after the transition to full alertness (larger $t_2$). PD neurons not always show such a region. Under deep anesthesia, PD neurons have an increased $t_1$ compared to the control group, with a temporal structure disruption at long scales. At full alertness PD neurons show an extended long-scale correlations window (regime III).

Our interpretation of these findings in the network context is as follows. Under deep anesthesia, the most marked difference between PD and control neurons is a significant prolongation of regime I of the temporal structure function. Since in this state external input is minimum, the observed differences between the two groups can be attributed to a differing small-range network structure, playing a stronger role in PD than in the control group. To corroborate this interpretation, we performed a simulation study. In this network, the small-range interactions were modeled by a diffusive coupling among the GPi cells. For the comparison of the PD vs. the control case, the identical network architecture was used. To recover the experimentally measured data, in addition to a stronger diffusion, also differing excitatory / inhibitory input levels were chosen (see below), in agreement with the hallmark of PD in the BG network [3]. In the simulation, each neuron had the form (Rulkov [11])

$$x_{i,n+1} = f(x_{i,n}, y_{i,n} + \beta_{i,n}) \quad (3)$$

$$y_{i,n} = y_{i,n} - \mu(x_{i,n}+1) + \mu\sigma + \mu\sigma_{i,n} \;, \quad (4)$$

where the index *n* indicates the iteration step, and where function *f* is given by

$$f(x_n, y) = \begin{cases} \alpha/(1-x_n) + y, & x_n \leq 0 \\ \alpha + y, & 0 < x_n < \alpha + y \quad \text{and} \quad x_{n-1} \leq 0 \\ -1, & x_n \geq \alpha + y \quad \text{or} \quad x_{n-1} > 0. \end{cases} \quad (5)$$

External input is modeled by

$$\sigma = \sigma_u + I, \quad (6)$$

where $\sigma_u$ represents the initial excitability of each isolated neuron. In total, 101 GPi neurons were implemented, aligned on a ring structure (Fig. 3). The Subthalamic (STN) and Striatal (Str) inputs to the GPi are modeled as excitatory and inhibitory inputs respectively, and the spatial distribution of both inputs is close to the available histological data [12]: Input to GPi is mediated by 101 STN axons, each of which sends collaterals to 10 neighboring cells using identical synaptic weights ($w_{\text{STN-GPi}}$=0.1). Str input to the GPi is mediated by 101 axons producing 10 collaterals each: one central connection to a GPi neuron with a high synaptic weight ($w_{\text{Str-GPi-I}}$=0.9) and 9 connections to adjacent cells with a lower weight ($w_{\text{Str-GPi-II}}$=0.01). Inputs were modeled by uniformly distributed random numbers from the unit interval, multiplied by the amplitude $A_e$, for excitatory input or by amplitude $A_i$, for inhibitory input, respectively. The distinction between anesthetized and alert conditis is modeled by changed input amplitudes. Whereas the anesthetized condition was modeled by $A_e$=1.5/25 and $A_i$=-1.2/-24.5, the alert condition was modeled by $A_e$=2/50 and $A_i$=-1.5/-48.5 (control / PD cases, respectively). In this way, the two conditions were characterized by slightly different $A_e/A_i$ ratios of 1.25/1.02 (anesthesia) and 1.33/1.03 (alertness) for the control / PD case, respectively. These values were chosen in order to reproduce best our experimental measurements, but they also have physiological justification (cf. [3] and references contained).

The coupling from each neuron $i$ to its neighbors is described by the following equations (cf. [11], where we set $\beta_e$ and $\sigma_e$ to 1):

$$\beta_{i,n} = g_{ji} \beta^e (x_{j,n} - x_{i,n}) \quad (7)$$

$$\sigma_{i,n} = g_{ji} \sigma^e (x_{j,n} - x_{i,n}). \quad (8)$$

The local dependence of the coupling on the neighbor order $j$ is implemented by

$$g_{ij} = \frac{D}{|(i-j)|^2}.\qquad(9)$$

For every neuron, the parameter values $\alpha=4.5$ and $\mu=0.001$ were used. To account for variability in initial neuronal excitation, $\sigma_u$ was drawn uniformly from [0.05, 0.15]. The value of $\alpha$ was chosen to reproduce our experimental burst length, and the slight variations in $\sigma_u$ helped including in the model the subtle transitions from a bursting to a spiking regime that were observed during our experiments (for a detailed description of the parameter space of Rulkov neurons, see ref. [11]; for details on the bursting behavior of our measured neurons, see Appendix). As was stated above, the main difference between PD and control cases was a different choice of diffusion strength ($D=0.3$ in the PD vs. $D=0.01$ in the control case), in addition to the different input levels implemented. Once a stationary regime had been achieved, the system was iterated for 180,000 time steps and spikes were extracted to obtain ISI time series. On these data, the same structure function analysis as for the experimental data was applied.

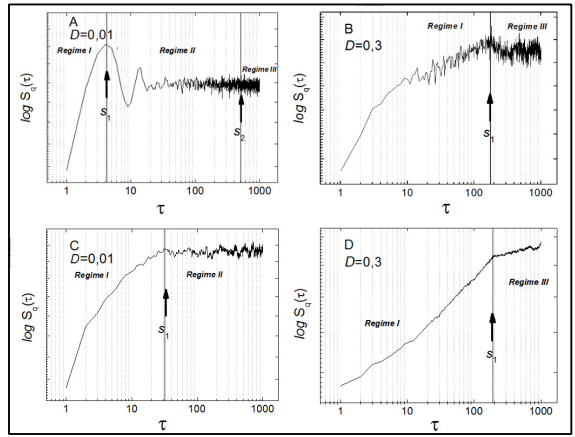

**Fig. 2) Modeling results** (characteristic examples, corresponding to Fig. 2). The main qualitative features measured are reproduced by changing the strength of diffusive coupling and the excitatory/inhibitory input levels (see text).

The simulations that are fully based on the available physiological data, show substantial agreement with our experiments (Fig. 2). The variants of behavior observed in our experiments were also encountered in our model. In the model, not all neurons evolve in exactly the same subset of the parameter space, because of the initial variability introduced. This fact explains the slight differences observed. Regarding the duration of regime I, in our simulations $t_1$ depends on two variables: the frequency of discharge has a direct relation with $t_1$, while the coupling strength $D$ shows an inverse relationship with it. In this context, we propose that the increased coupling strength may act as an adaptive response to the primarily increased frequency in the parkinsonian GPi. From this, our conclusion is that diffusive coupling is a key feature shaping the temporal structure of the neuronal discharge, and that its pathologic increase might lead to the abnormalities observed in PD. Moreover, we propose that this increase could explain the excessive correlation found in the BG [13-17]. Experimental evidence from other works supports this interpretation: under PD, the coupling between adjacent Striatal cells has been shown to be significantly increased [18-19]; a similar effect might also take place in the GPi.

At all alertness levels, the rate-coding window is extremely shortened for PD neurons. Under anesthesia, small-range dynamics dominates the neuronal firing in PD, while at full alertness, almost all the PD dynamics is captured by long-term correlations. This makes the PD system extremely sensitive to noise, since all fluctuations of the environmental input on temporal scales longer than a characteristic time (our time scale $t_2$) trigger long-time correlations in the system. In the control group, on the contrary, the transition to full alertness is characterized by an extension of the rate-coding window. This is a cell-correlate of de-synchronizing neural populations reported from electroencephalographic studies during this transition. In this way, healthy neurons are robust to noise over larger time- and frequency-scales than PD neurons, and this difference is further augmented towards full alertness.

Our findings are relevant to the application of deep brain stimulation (DBS) therapy. DBS consists in chronically stimulating a selected BG target with a train of high frequency electric pulses, which may successfully alleviate the symptoms of PD and also of other neurological conditions, such as dystonia, Tourette-syndrome and essential tremor. However, the mechanism of action of DBS has remained to be unknown, and therefore we currently lack a rational basis for the selection of the optimal stimulation parameters (train frequency, pulse amplitude and width) and the specific stimulation target (GPi and STN are preferred ones). On this basis, the outcome of DBS therapy is hard to predict in individual patients. Understanding the mechanisms that underlie the coding of information in the BG and how they are altered by PD is a key issue for understanding how PD neurons respond to DBS. For the first time we showed that several coding-schemes with characteristic time-scales are present in the BG, and that they are altered in PD due to excessively strong diffusive coupling (for experimental evidence from human PD patients at full alertness, cf. [20]). We suggest that successful DBS not only breaks the pathologic correlations between PD neurons, but also restores the rate-coding window at the proper time-scale, which in turn re-establishes a robust coding of information in the BG. We expect that our approach to PD will be helpful to further explore presently not well understood aspects of PD and related disorders.

**Acknowledgments.** We acknowledge the support by the technical personnel at the laboratories of the Center for Applied Neurological Research, Fleni Institute, Buenos Aires, Argentina.

Appendix

Supplementary material to: A hierarchical coding-window model of Parkinson's disease

**Authors**: Daniela Andres [*,1,2,3], Florian Gomez[1], Daniel Cerquetti [2], Marcelo Merello [2], Ruedi Stoop [1]

* **Corresponding author**: dandres@ini.uzh.ch

**Affiliations**:

1. Institute of Neuroinformatics, University and ETH Zurich, Zurich, Switzerland

2. Institute for Neurological Research Raul Carrea, Fleni Institute, Movement Disorders Section, Buenos Aires, Argentine

3. Society in Science, The Branco-Weiss Fellowship, administered by ETH, Zurich, Switzerland

Background

The surgical treatment of Parkinson's disease (PD), including Deep Brain Stimulation (DBS) therapy, has been successfully used in selected groups of patients for many decades [1-7]. In this treatment, the implantation of DBS electrodes is performed under stereotactic functional neurosurgery and the utilization of microrecording of the neuronal activity along the surgical tract is usually considered the gold standard procedure for target identification for the placement of the stimulation electrodes [8-11]. Additionally, this technique allows the recording of neuronal extracellular activity, from which later single units can be isolated and their data can be used for a detailed analysis [12-13].

Regarding the application, indication and mechanisms of action of DBS [14-18], many questions still remain open. In previous works we have emphasized the necessity of analyzing data recorded from human brains in vivo, both for the understanding of the pathophysiology of PD and for technological improvements of the procedure. However, the interpretation of results based on this data is difficult to achieve by several facts. As a first issue, as the neuronal data is obtained from human brains, recording length and recording site localization precision are limited by the surgical risk for the patient. This risk increases with total surgery duration, so that some surgery centers will generally accept a prolongation of the surgery of no longer than 30-45 minutes, which usually allows the recording of segments of 2-3 minutes of neuronal activity only. This strongly limits the application of time series tools that generally critically depend on the data length. Moreover, as the primary purpose of microelectrode recordings during DBS surgery is target identification, the microelectrode is generally kept moving, to find the best implantation target. When the electrode is not moving, the patient can be asked to perform tasks, to evaluate the neuronal response or to conduct routine surgical controls. Such conditions, even when following strict recording protocols, cannot be regarded as stationary. Finally, microelectrode recording of neuronal activity is performed in most cases under local anesthesia [19-21].

This implies that the patient is alert during the procedure, with the degree of alertness possibly affecting neuronal activity. In this way, measured properties of the neuronal discharge of PD patients might depend on the alertness level as well as on the severity of the disease.

To explore this dilemma, a control group by healthy human subjects is unavailable, so that the use of preclinical models of PD based on animal models emerges as the only possible approach. Animal models of movement disorders have been developed involving different species, ranging from invertebrate models of Drosophila Melanogaster (useful particularly for genetic studies), to primate models of PD [22]. Rodent models of PD include mice and rats treated with rotenone, 6OHDA or MPTP [23]. We use the well known 6OHDA-striatal-lesion model of parkinsonism in the rat. This model has shown not only to be cost-effective, but it has also been demonstrated to be a suitable model for the study of pathophysiological properties of the Basal Ganglia (BG) under the effect of dopamine-deprivation [24]. We use this model to study the effect of the level of consciousness on the neuronal activity of the GPi in the healthy and dopamine-depleted rat BG, independently of explicit visual and auditory sensory stimuli, and motor influence.

Experimental design

There are technical challenges associated to the recording of neuronal activity of deep brain structures. In particular, the discrimination between the effect of the level of alertness or consciousness and that of sensory stimuli or motor activity can be difficult, since it requires stereotactic exploration of deep brain structures in the head-restrained, alert and relaxed animal. Recently, studies in alert, head-restrained animals were conducted with devices that required training of the animals for its utilization [25-27]. Other studies have attempted to analyze the effect of the level of alertness on the neuronal behavior of different neural systems, sometimes making use of implanted electrodes that allow recording neuronal activity in freely moving animals [28-30]. Regarding the human BG in PD patients, contradictory results were reported. While some authors found no statistical difference between alertness and deep anesthesia in the neuronal discharge of the Subthalamic Nucleus (STN) and Substantia Nigra pars reticulata (SNr) [31-33], Raz et al. reported a significant decrease in the discharge rate of STN neurons under propofol effects [34].

We designed a dedicated device that required no previous training of the animals, which allowed us taking long recordings of neuronal activity *in vivo*, for periods of time lasting up to several hours. In this way, we achieved the recording of single neurons' during the transition from deep anesthesia to full alertness in a deep brain structure in the alert, relaxed, head-restrained animal. All the experiments were conducted under identical conditions of environmental silence and with the eyes of the animals covered, guaranteeing the absence of both auditory and visual sensory input, as well as motor activity at the time of the recordings. Following this procedure we were able to obtain stationary conditions and isolate the effect of an alert state of consciousness at a cell level, i.e. on the neuronal discharge of the GPi. With the exception of few carefully selected cases, this kind of study cannot be performed in human patients (see for example Ref. 33), in whom the microelectrode-recording of neuronal activity usually takes place under local anesthesia only.

Ethics statement

All animal experiments and procedures were conducted with adherence to the norms of the Basel Declaration [35]. The experimental protocol was revised and approved by our local ethics committee CEIB, Buenos Aires, Argentine. All experiments took place at the authorized laboratories of our centre, and adherence to the Basel Declaration standards were monitored by our research staff. During the time previous and between experiments, animals were housed in racks with optimal temperature, pressure and air humidity regulation under an inverted 12 hours light cycle, with water and food available *ad libitum*. To minimize animal suffering, optimal anesthetic and analgesic medication were used as described below. Euthanasia was conducted using a high dose of meperidine, an opioid suitable for that purpose, guaranteeing the absence of animal suffering during the procedure. The number of animals used in the experiments was the minimum considered necessary to achieve sound conclusions. See APPENDIX A: veterinary letter of support.

Animal model

Adult male and female Sprague-Dawley rats weighting 250-350 gr. were randomly divided in 3 groups: 6-hydroxydopamine-lesioned group (6OHDA, n=16), Sham-lesioned group (n=10), and not-lesioned group (n=10). Animals within the 6OHDA group were lesioned unilaterally following the partial-lesion model originally described by Sauer and Oertel [36]. A total dose of 20 µg of 6OHDA diluted in 4 µl of saline solution supplemented with 0.2 mg/ml of ascorbic acid was injected at a rate of 1 µl/min in the Striatum (coordinates: anterio-posterior, +1.0; lateral, +3.0; depth, -4.5) using a Hamilton microsyringe. After waiting five minutes from the time the injection was completed, the syringe was withdrawn from the lesioning site at a rate of 1 mm / min. Animals in the Sham group underwent the same surgical procedures as those in the 6OHDA group, but were injected only with the vehicle (ascorbic acid solution, at the same concentration mentioned above). All 6OHDA and Sham lesions were placed in the left hemisphere. During the surgery, temperature maintenance was cared for. The surgery was conducted with the aid of a stereotactic frame (Small Animal Stereotaxic Instrument, LS900, David Kopf Instruments, Tujunga, CA, USA) and coordinates were assessed by use of the Paxinos & Watson Atlas [37]. Animals were placed in the frame and reference points were defined. In the horizontal plane the skull point bregma was taken as the reference point. Its position was assessed with the help of an optical microscope. The cortical surface was considered the reference point along the vertical axis, and its position was defined with the aid of electrical means.

Behavioral evaluation

Between 21 and 28 days after the lesion procedure, animals were evaluated using the cylinder test [38], which serves the purpose of quantifying asymmetry in motor behavior. The animals were placed in a transparent cylinder for 5 minutes and left to explore freely, and only weight supporting touches of the wall were counted, according to the criteria described by Lundblad et al. [39] for the cylinder test. All tests were video-recorded. Fig. 1 shows the results of the behavioral evaluation for the three groups analyzed. We recorded a total of 22.17 ± 9.96 (mean ± SD) touches / test. The left (negative) to right (positive) bias was calculated as the percentage difference between left and right touches over total number of touches. Animals in the 6OHDA group showed a marked asymmetry towards the side ipsilateral to the lesion, reflected by a left bias in the use of the front-limb for weight supporting touches of the wall. On the contrary, both control groups (the Sham and the not-lesioned groups) showed a right bias. A statistical difference with a p-value < 0.01 was obtained between the 6OHDA and both the Sham and the not-lesioned groups, while no significant difference was observed between the last two. The absence of overlap seen between the 6OHDA-lesioned and both control groups evidenced the successful implementation of the 6OHDA- vs. Sham-lesion model.

Anesthesia, analgesia and antibiotic medication

Three complementary drugs were used for achieving anesthesia and analgesia in our study: chloral hydrate, tramadol and lidocaine. Animals were injected with a 300 mg/kg intraperitoneal dose of chloral hydrate (at a concentration of 50mg/ml) used as anesthetic. At this dose and concentration of chloral hydrate, a mortality rate of 0% in adult rats has been reported, while sufficiently deep anesthesia for surgical procedures is achieved [40]. Anesthesia can be defined as the concomitant presence of unconsciousness, analgesia and muscle relaxation [41]. In current approaches to anesthesia, these effects are usually obtained with combinations of multiple drugs, since this allows using lower doses and therefore minimizing morbimortality [42]. Chloral hydrate is a well known sedative with potent hypnotic effects, widely used not only in veterinary medicine but also in pediatric and neonatal medicine [43-45]. However, it does not have important analgesic effects. In the current protocol, analgesia was achieved with a 4 mg/kg dose of intraperitoneal tramadol. Tramadol is a drug commonly used in veterinary medicine, which has combined mechanisms of action, a wide safety rank, few side effects and has proven effective for managing surgical pain [46-49]. Tramadol has been shown to have an analgesic potency similar to meperidine and morphine for treating pain of different origins, including surgical pain [50-51]. In our protocol, the tramadol dose was repeated between 12 and 24 hours after the surgery to maintain analgesia, and therefore it was used both as preemptive analgesic and as postoperative medication. Local anesthesia (lidocaine) was used at the incision and at contact points. The eyes of the animals were protected from corneal drying with ophthalmic solution drops. Antibiotic prophylaxis was administered in the form of a single 10 mg/kg dose of intramuscular cefazoline previous to the surgery.

The choice of the mentioned drug profile responded to particular issues related to the disease model implemented (6OHDA-lesion model of Parkinson's disease). Other drug options in laboratory animals include dissociative anesthetics (in particular ketamine and combinations of this drug, for example ketamine-xylazine), barbiturates and inhalant anesthetics (halothane, isoflurane, sevoflurane, among others). Ketamine-xylazine has been reported to yield neuroprotective effects over the central nervous system (CNS), and it is currently questioned to what degree it interferes with the 6OHDA Parkinson's disease model, which made it unsuitable for the present study [52]. In the case of isoflurane, it has been shown to induce apoptosis in the CNS [53-54], which could also have potentially affected the implementation of the 6OHDA-model. Since other inhalant anesthetics haven't been analyzed regarding this effect, it is safer to avoid this drug family. Halothane might have been an alternative, but it is a highly hypotensive and arrhythmogenic drug, not necessarily safe in small animals [55]. Barbiturates, on the other hand, are also known to have very narrow safety margins [56].

The anesthetic and analgesic medication used was the same for the lesion- and the recording surgeries, and the stereotactic procedure was repeated as well.

Assessment of anesthesia depth

During the neuronal activity recording surgery, the state of consciousness was characterized periodically (every 10-12 min) by evaluating the tail-pinch reflex with the application of a standardized non-painful stimulus. Methods to assess the state of consciousness under anesthesia in animals have been widely discussed [57-58]. We defined the following levels of alertness. Level 1: deep anesthesia, level 2: mild alertness, level 3: full alertness. At level 1, animals did not present a positive paw withdrawal reflex, cutaneous reflex or tail-pinch reflex. Level 2 was defined as the first time of appearance of a positive tail-pinch reflex, and at level 3 the animals showed a strong response to tail-pinch stimulation, either withdrawing both paws and / or energetically contracting abdominal muscles. Since the evaluation of reflexes is subjective, all the evaluations were conducted by the same person, to avoid inter-personal variation. At the end of the recording surgery we confirmed the wellness and alertness level of the animals by letting them explore freely through the laboratory.

Recording of neuronal activity

After completed motor evaluation, animals went through stereotactic surgery with the objective of registering spontaneous activity of the *medial Globus Pallidus* or *Globus Pallidus pars interna* (mGP or GPi). This nucleus corresponds to the structure that was previously called entopeduncular nucleus. In the present work we adhere to the nomenclature proposed by the latest edition of Paxinos & Watson's Atlas and refer to this structure as mGP or GPi. The time between the lesion and the recording surgery was between 21 and 28 days. This amount of time has been proven to account for a significant level of dopamine-cells and dopamine-content loss in comparison to the contralateral, not-lesioned side of the brain [36]. In our work we show with a behavioral test that motor deficits followed the injection of 6OHDA but not the Sham-lesion, which makes the model implemented suitable for the study of pathophysiological features of PD. Similar times were allowed between both surgeries for the Sham-group. Recording coordinates fell within the limits defined as the GPi by Paxinos & Watson's Atlas, and are shown if Fig. 2. Following anesthesia, animals were placed in a specially designed restraining device, which was built ad-hoc with semi-rigid plastic and covered in the inside with a soft and high quality thermal insulator. The device's purpose was not to keep the animals firmly restrained if they spontaneously moved, as it was only loosely bound. On the contrary, it served the purpose of minimizing discomfort helping animals to stay calm during the surgery. During the whole recording time, the animals did not make any spontaneous movements. If the animals didn't relax but, on the contrary, attempted to move during the experiment (which happened only in few cases) we considered that an end-point for the recording-surgery. During the whole surgical procedure, the eyes of the animals were covered and all surgeries were conducted in identical conditions of environmental silence. Neuronal activity uninterruptedly during the awakening process and for long periods of time, obtaining up to three hours of recording of the same neuron.

Neuronal recordings were obtained using glass-insulated platinum/iridium (Pt/Ir 80/20%) microelectrodes with nominal impedance of 0.8–1.2 megohms (mTSPBN-LX1, FHC Inc, Bowdoin, ME, USA). Signals were amplified, conditioned and monitored with an analog oscilloscope, digitized with a dedicated acquisition system (1401plus, CED) and saved in a PC running Spike 2.0 software. The sampling rate was 20 kHz and total amplification including probe was x10,000, checked with a built-in calibration signal of 1mV p-p at the beginning of each experiment. Fig. 3 shows sample recording segments of bursting and tonic-firing activity. The mean length time of the recordings was 58.98 ± 30.82 min. No statistically significant differences were observed between the Sham and not-lesioned animals for any of the characteristics analyzed in the present work, which will therefore be reported as a single group and referred to as control group.

Data analysis

Signals were processed off-line. Spikes were extracted and classified using the algorithm developed by Quian Quiroga [59]. Single units were used to construct interspike intervals (ISI) time series. In order to guarantee stationary conditions, thirty seconds of recording following the application of a tail-stimulus were discarded from each time series. To be able to characterize different alertness levels unaffected by the proper transition between them, the time separation between the recording segments analyzed was maximized. For alertness levels 1 (deep anesthesia) and 3 (full alertness), the first and the last segments of the recording were selected, respectively. In this way, a time separation of 22.52 ± 3.79 min (mean ± SEM) was obtained between successive recording segments. Although the transition segments were not used for the analysis, the recordings were uninterrupted during the whole awakening process with the objective of guaranteeing that the same neuron was being recorded at different alertness levels.

Bursting and statistical properties

We studied the statistical properties of the ISI time series (mean, median, mode, percentiles, standard deviation, skewness and kurtosis), and also analyzed the frequency of discharge and bursting activity of each neuronal group. Statistical comparisons between groups were performed using the non-parametric Kolmogorov-Smirnoff test and two-sample t-test, and a p-value<0.05 was considered significant. We did not find differences between the groups studied in the rate of discharge under the effect of anesthesia, but as the level of alertness increased the rate of discharge of the neurons in the 6OHDA group was incremented while the opposite occurred for the control group (Fig. 4). Although these changes were not statistically significant within groups, a significant difference was found in the alert state (anesthesia level 3) between parkinsonian and control neurons, with a higher discharge rate in the 6OHDA group. The second up to fourth moments of the probability distributions of ISI did not show significant differences between groups.

The bursting activity was quantified by counting the percentage of spikes that triggered a burst (Burst Triggering Spikes, BTS). We based our detection of bursts on the rank-surprise algorithm [60] and defined the BTS-index as the percentage of bursts over the total number of spikes. As a value limit for the largest ISI to be considered to be part of a burst we used the p75 of the ISIs distribution. The minimum surprise value accepted was $\alpha = -\log(0.01)$. We did not find any statistically significant differences between the 6OHDA and control groups. However, BTS-index diminished consistently with increasing level of alertness, showing that the bursting activity of single neurons can vary adaptively depending on the anesthesia depth. This suggests that bursting behavior should be considered as an emergent property of the state of the neuronal system rather than an intrinsic electrophysiological property of individual GPi neurons.

Figures and tables

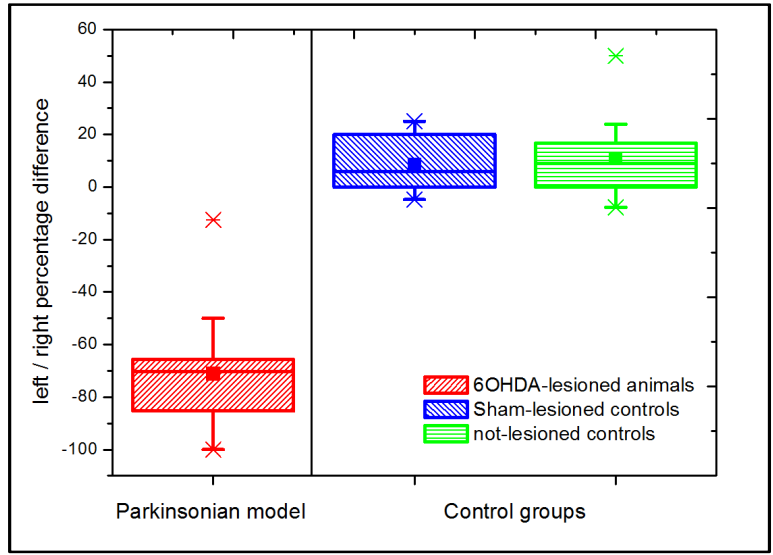

**Figure 1. Cylinder test results.** Percentage differences in weight supporting wall touches with the left (negative) vs. right (positive) paw are shown. Statistically significant differences were observed between the parkinsonian model (n = 16) and both the Sham-lesioned (n = 10) and the not-lesioned (n = 10) control groups (p < 0.01). This not-overlapping behavioral differences between parkinsonian (6OHDA-lesioned) animals and both control groups follow the correct implementation of the model. The boxes edges represent percentile 75 (p75) and percentile 25 (p25) respectively, the inner line indicates the median and whiskers account for p75+1.5(p75 - p25) and p25-1.5(p75-p25) respectively. The mean is represented by the inner, solid box. Crosses indicate outliers.

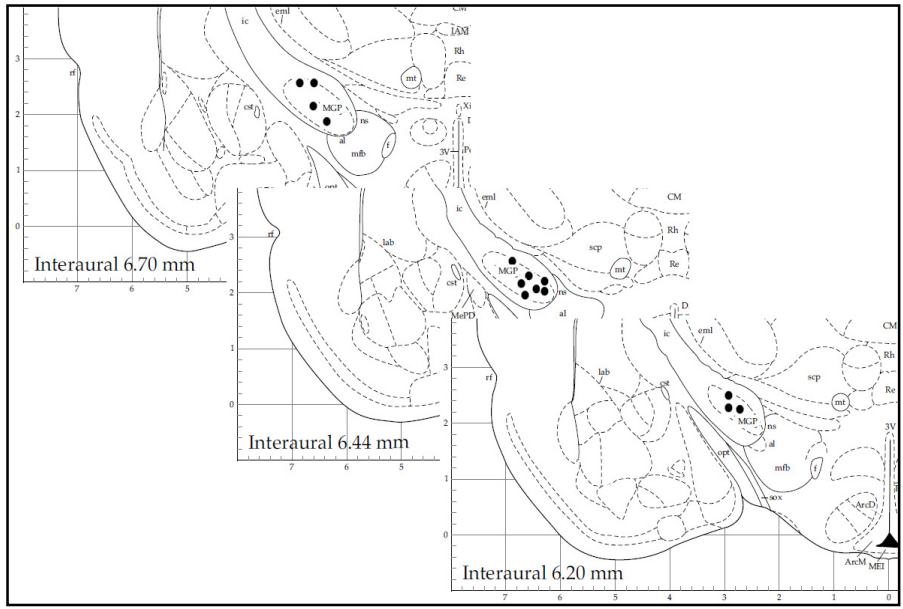

**Figure 2. Stereotactic coordinates of recording sites for 15 sample neurons.**

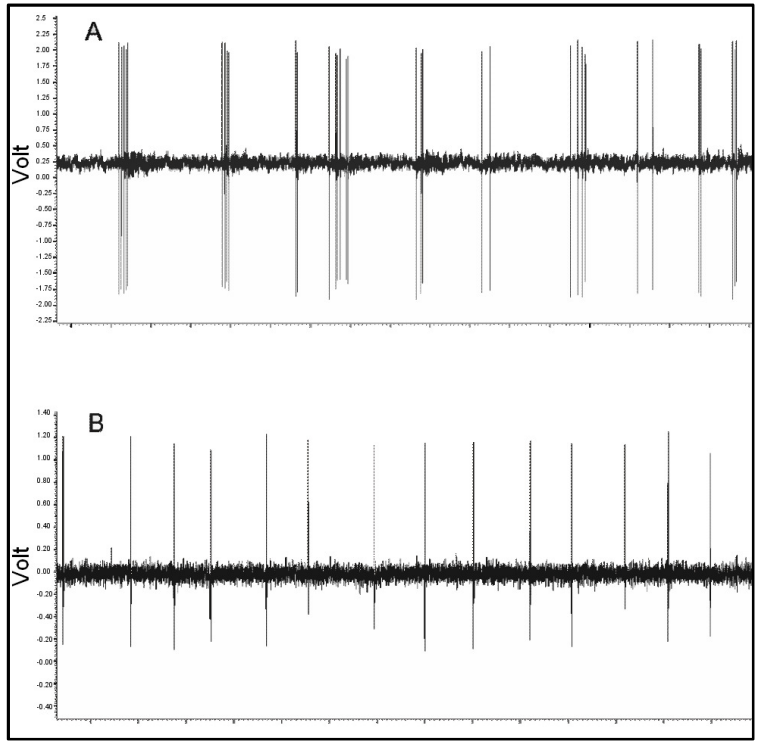

**Figure 3. Sample bursting (A) and non-bursting (B) segments of the neuronal activity recorded.**

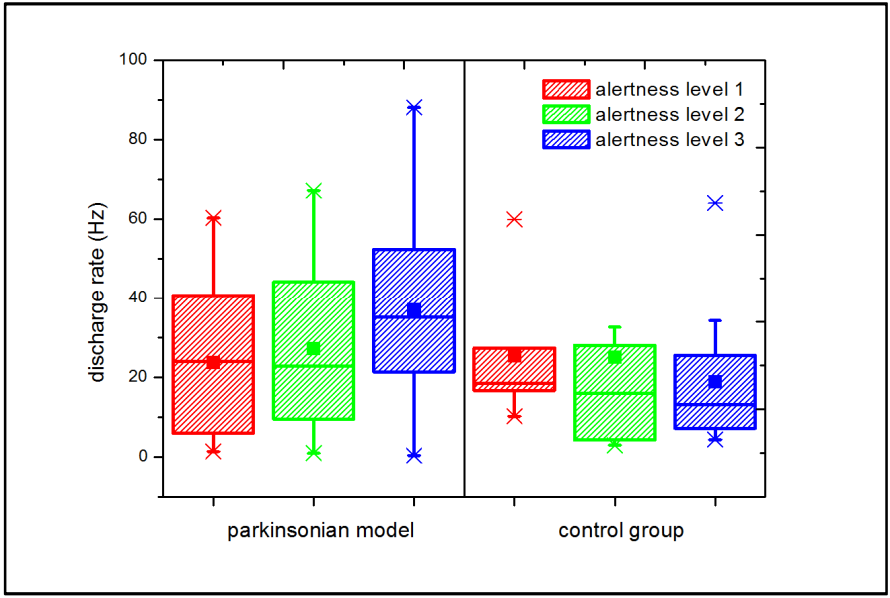

**Figure 4. Rate of discharge at different levels of alertness.** An increment in the discharge rate is observed in the parkinsonian neurons following increasing levels of alertness. A statistically significant difference was observed between the parkinsonian (n = 17) and control (n = 21) neurons at alertness level 3, ($p<0.05$), showing that this difference depends on the input received by the neural system. Upper and lower edges of the boxes represent percentile 75 (p75) and percentile 25 (p25) respectively, the inner line indicates the median, upper and lower whiskers account for p75+1.5(p75-p25) and p25-1.5(p75-p25) respectively. The mean is represented by the inner, solid box. Crosses indicate outliers.

APPENDIX A) Veterinary letter of support.

Buenos Aires, June 12th, 2013

To whom it may concern,

Hereby I state that I have provided professional advice as an independent Veterinary Medic consultant to the staff of the Experimental Parkinson's Laboratory, Center for Applied Neurological Research, Fleni Institute, Escobar, Buenos Aires, Argentine, during the design and implementation of the protocol labeled "Microelectrode recording of neuronal activity from deep brain structures in a group of 6-OHDA-lesioned and a control group of adult Sprague-Dawley rats, under chloral-hydrate deep anesthesia and in alert, head-restrained conditions".

Based on professional and clinical expertise, I considered the types, doses and administration routes of drugs chosen for the surgeries conducted in the frame of the mentioned protocol to be adequate in order to achieve correct surgical anesthesia and analgesia, and also euthanasia, in the group of animals used for the protocol.

The drugs administered to the animals were the following:

- Anesthesia: chloral-hydrate, 300 mg/kg, intraperitoneal injection, at a concentration of 50 mg/ml.
- Analgesia: tramadol, 4 mg/kg, intraperitoneal injection. One dose previous to the surgery and a second one, 12 - 24 hs. postoperatively.
- Antibiotic profilaxis: single dose of cefazoline, 10 mg/kg, intramuscular injection, previous to the surgery.
- Local anesthesia: lidocaine as needed, at incision and contact points.
- Euthanasia: high dose of meperidine, intraperitoneal injection.

Your sincerely,

Virna Koch, MP 10944 (Argentine)

Dr. of Veterinary Medicine